\documentclass[aps,prl,twocolumn,groupedaddress,showpacs]{revtex4}
\usepackage{graphicx}
\usepackage{psfrag}
\def\be{\begin{equation}}
\def\ee{\end{equation}}

\begin{document}
\date{\today}

\title{Core-halo distribution in the Hamiltonian Mean-Field Model}

\author{Renato Pakter}
\author{Yan Levin}
\affiliation{Instituto de F\'{\i}sica, Universidade Federal do Rio Grande do Sul, Caixa Postal 15051, CEP 91501-970,
Porto Alegre, RS, Brazil}

\begin{abstract}

We study a paradigmatic  system with long-range interactions:
the Hamiltonian Mean-Field Model (HMF).
It is shown that in the thermodynamic limit this model does not relax to the usual equilibrium 
Maxwell-Boltzmann distribution.
Instead, the final stationary state has a peculiar core-halo structure.  In the thermodynamic limit, HMF 
is neither ergodic nor mixing.  Nevertheless, we find that using dynamical properties of Hamiltonian systems,
it is possible to quantitatively predict both the spin distribution
and the velocity distribution functions in the final stationary state, without any adjustable parameters. 
We also show that HMF undergoes a non-equilibrium first-order phase transition between  paramagnetic and ferromagnetic states.

\end{abstract}


\maketitle

Since the early  work of Clausius, Boltzmann, and Gibbs it has been known that for particles interacting
through short-range potentials, the final stationary state 
reached by a system corresponds to the thermodynamic equilibrium~\cite{Gibbs}.  Although no exact 
proof exists, in practice it is found that non-integrable systems with 
fixed energy and number of particles (microcanonical ensemble) always relax to a unique
stationary state which only depends on the global conserved quantities:  energy, momentum, and angular momentum.
The equilibrium state does not depend on the specifics of the initial particle distribution.  The situation is very
different for systems in which particles interact through long-ranged unscreened potentials.  This is 
the case for gravitational systems and confined one component plasmas~\cite{Levinprl,TaLe10}. 
For these systems, in the thermodynamic limit, 
collision duration
time diverges, and the  thermodynamic equilibrium is 
never reached~\cite{Kav2007}.  Instead, as time $t \rightarrow \infty$,
these systems become trapped in a stationary state characterized by a 
broken ergodicity~\cite{Barre2001,Muka2005,Campa09}. Unlike the thermodynamic equilibrium, the stationary state 
depends explicitly on the initial particle distribution.
Over the last $50$ years, there has been a great effort to predict the final
stationary state without having to  explicitly solve the $N$-body dynamics or  
the collisionless Boltzmann (Vlasov) equation. Qualitatively, it has been observed that for
many different systems the non-equilibrium stationary state has a peculiar core-halo shape.  Recently, an
ansatz solution to the Vlasov equation has been  proposed which allowed us to explicitly calculate 
the core-halo distribution function
for confined plasmas and self-gravitating systems~\cite{Levinprl,TaLe10}.
In this Letter we will show that an ansatz solution is also possible for the  
HMF model.  The theory proposed allows us also to locate the non-equilibrium para-to-ferromagnetic
phase transition, which  earlier theories 
incorrectly predicted to be of second order~\cite{AnFa07}. 
All the results are compared with the molecular dynamics simulations
performed using symplectic integrator, and are found to be in excellent agreement.

The HMF model consists of $N$, $XY$ interacting  spins,  whose dynamics
is governed by the Hamiltonian
\be
H=\sum_{i=1}^N {p_i^2\over 2}+{1\over 2 N}\sum _{i,j=1}^N [1-\cos(\theta_i-\theta_j)],
\ee
where angle $\theta_i$ is the orientation of the $i$th spin 
and $p_i$ is its conjugate momentum~\cite{AnFa07,AnFa07b,AnCa07}.
The {\it macroscopic} behavior of the system is characterized by the magnetization vector
${\bf M}=(M_x,M_y)$, where $M_x\equiv \langle \cos\theta \rangle$, 
$M_y\equiv \langle \sin\theta \rangle$, and $\langle\cdots\rangle$ stands for the average
over all particles. The modulus, $M=|{\bf M}|$ serves as the order parameter which measures the
coherence of the spin angular distribution: for $M=0$ we have a completely
incoherent state, whereas for finite $M$ there is some degree of coherence. 
Hamilton's dynamic equations for each spin can be expressed in terms of 
the total magnetization and take a particularly simple form $\ddot\theta_i=F(\theta_i)$, where the
force on each spin is $F(\theta_i)=-M_x\sin\theta_i+M_y\cos \theta_i$.  
%\be
%\ddot\theta_i=-M_x\sin\theta_i+M_y\cos \theta_i.
%\label{evol}
%\ee
The average energy per particle is $u=H/N= p^2/2+(1-M^2)/2$.
%\be
%u={H\over N}={\langle p^2\rangle\over 2}+{1-M^2\over 2}\,.
%\label{energyu}
%\ee 
Since the 
Hamiltonian does not explicitly depend
on time,  $u$ 
is conserved along the temporal evolution. For simplicity  
we will consider initial distributions of ``water-bag'' form in the $(\theta,p)$ 
reduced phase-space ($\mu$-space). 
Without loss of
generality, we choose a frame of reference where $\langle \theta \rangle =0$ and 
$\langle p \rangle =0$. The one-particle initial distribution function then reads
\be
f_0(\theta,p)={1\over 4\theta_0 p_0} \Theta (\theta_0-|\theta|)\, \Theta (p_0-|p|),
\label{f0}
\ee
where $\Theta$ is the Heaviside step function, and $\theta_0$ and $p_0$ are the maximum absolute values of the angle and 
the momentum, respectively.
Such initial distributions
are characterized by $M_x=M_0$, $M_y=0$, and $u=p_0^2/6+(1-M_0^2)/2$, where 
$M_0= \sin(\theta_0)/\theta_0$ is the initial magnetization. Because of the symmetry of $f_0$ with
respect to $\theta=0$, in
the thermodynamic limit $M_y=0$ throughout the evolution, so that the macroscopic dynamics is
completely determined by $M_x(t)$.

As the system evolves, the particle distribution in the phase-space changes and 
eventually reaches a stationary state or a limit cycle.
If $N$ is finite, the  stationary  state will be described by the equilibrium
Maxwell-Boltzmann (MB) distribution.  In the thermodynamic limit  $N\rightarrow \infty$, 
however, the system becomes trapped
in a non-ergodic non-mixing state, the life time of which diverges with the number of particles. 
In this  limit, 
the dynamical evolution of one-particle distribution function $f(\theta ,p,t)$
is governed {\it exactly} by the collisionless Boltzmann (Vlasov) equation~\cite{Br77}, 
\be
\frac{\partial f}{\partial t}+p \frac{\partial f}{\partial \theta}+F(\theta) \frac{\partial f}{ \partial p}=0 \,.
\label{vlasov}
\ee
The left-hand side of this equation 
is just the convective derivative of the one particle distribution function.   
Therefore, a collisionless Hamiltonian system evolves  over the phase space as an incompressible fluid. 
Furthermore, the incompressibility implies that during the  
temporal evolution the phase-space density can never exceeds the maximum of the 
initial distribution function.

Although the MB 
distribution is also a stationary solution of the Vlasov equation,  unlike for
Boltzmann equation, it is not a global attractor of the
dynamics, so that  an arbitrary  initial distribution will not evolve to the
MB equilibrium. The collisionless relaxation described by the Vlasov equation  
for systems with long-range interactions is, therefore,
much more complex than the collisional relaxation governed by the usual Boltzmann equation
for systems with short-range forces. 

Vlasov equation is time reversible. Thus, on a fine-grained scale, temporal evolution never ends.  
There is no fine-grained attractor for the Vlasov dynamics.  In practice, however, 
one can never have infinite resolution, and there is a finite maximum precision that one can reach in any experiment
or a numerical simulation.  It is on this coarse-grained scale that it {\it appears} that the evolution has
reached a steady state.  Unlike for Boltzmann equation, however, the stationary coarse-grained
distribution function depends explicitly on the initial condition.  

%Vlasov dynamics is very peculiar in that it
%has an infinite number of conserved quantities   --- called the Casimir invariants --- 
%which are preserved by the collisionless Hamiltonian flow.
%The one-particle distribution function evolves in the phase space as an incompressible fluid.
%Therefore,  temporal evolution can never lead to a phase-space density that exceeds maximum of the 
%initial distribution function.

Recently it has been observed that for systems with long-range interactions, 
such as self-gravitating clusters and plasmas \cite{Levinprl,TaLe10}, the final stationary state has a peculiar 
core-halo structure.  The mechanism of  core-halo formation is 
very similar to the process of evaporative cooling.  
As the dynamics evolves,  macroscopic propagating density waves are formed.  Some particles
enter in resonance with the macroscopic oscillations 
gaining {\it large} amount of energy at the
expense of the collective motion.  This is similar to the mechanism of 
Landau-damping well known in plasma physics~\cite{La46}.
Resonant  particles can gain sufficient energy to 
reach high energy states, thus forming a diffuse halo. On the other hand, the loss of energy dampens the macroscopic oscillations,
so that the leftover particles become condensed into the low energy states, resulting in a dense core.
However, because of the 
incompressibility of the Vlasov dynamics, 
the core cannot completely freeze -- i.e., collapse to the minimum of the potential
energy. Instead, the distribution function of the core particles progressively approaches the maximum phase space density
allowed by the initial distribution --- all the low energy states become fully occupied by the core 
particles. Although the HMF model appears to be very different from either self-gravitating clusters or confined plasmas,
we find that its dynamical evolution follows exactly the same scenario as described above. 
%This suggests
%that there is some significant {\it universality} to collisionless relaxation.  
%of non-equilibrium systems with long-range interactions that 
%have a universal core-halo 
%distribution function, that is a  coarse-grained attractor of their dynamical evolution.  
   
In the case of the HMF, the oscillations of the magnetization
$M$ play the role of collective oscillations which drive some spins to higher
energy states, leading to a halo formation. 
The macroscopic oscillations of $M$ are significantly damped in one or two periods of oscillation.
The extent of the halo is determined on the same time scale.
As a consequence of the conservation of the total energy, 
the remaining spins populate lower and lower energy states, until all of them become fully occupied up
to the maximum phase space density $\eta_0=1/4\theta_0p_0$.
In the final stationary state, the core distribution function is the same as that of a 
fully degenerate Fermi gas of spin-degeneracy $\eta_0$.  The core distribution  extends up to the Fermi energy
$\varepsilon_F$. The value of $\varepsilon_F$ is yet unknown, and must be determined 
self-consistently.
We propose an ansatz for the core-halo distribution that  describes the final (coarse-grained) stationary
state reached by the HMF model at the end of its dynamical evolution:
\be
f_s(\theta,p)=\eta_0 \left [\Theta(\varepsilon_F-\varepsilon)
+\chi \Theta(\varepsilon_h-\varepsilon) \Theta(\varepsilon-\varepsilon_F)\right ],
\label{fs}
\ee
where $\varepsilon(\theta,p,M_s)=p^2/2+1-M_s\cos\theta$ is the single-spin energy, $\chi$ 
is the ratio between the halo and the core phase-space  densities, $M_s$ is the stationary value
of magnetization,
and $\varepsilon_h$ is the maximum energy of the halo spins. The energy
$\varepsilon_h$ is determined  from the short-time dynamics of spins driven by the
oscillations of the magnetization. To estimate this value we need an independent 
equation that  
describes the dynamical evolution of magnetization. We proceed as follows:
taking the second derivative of $M_x$
we obtain
$\ddot M_x=M_x\langle \sin ^2\theta\rangle-\langle p^2\cos\theta \rangle$.  
This equation requires the knowledge of the temporal evolution of $\langle \sin ^2\theta\rangle$
and $\langle p^2\cos\theta \rangle$, which leads to an infinite hierarchy of
equations. To truncate the hierarchy, we assume that for short times $\langle \sin ^2\theta\rangle=1/2$
and $\langle p^2\cos\theta \rangle=\langle p^2\rangle\langle \cos\theta \rangle =(2 u-1+M_x^2)M_x$,
where use has been made of the conservation of energy, together with the
condition $M=M_x$. We  then find a dynamical equation satisfied by the magnetization,
\be
\ddot M_x=-M_x\left(2u+M_x^2-\frac{3}{2}\right).
\label{mx}
\ee
This equation can be integrated numerically to provide the temporal evolution of $M_x(t)$. 
Since Eq.~(\ref{mx}) was derived neglecting the correlations between angles and momentums,
its validity extends only to very short times. However, the maximum energy of  the halo  is also
determined by the very short-time dynamics.  Thus, Eq.~(\ref{mx}), should be sufficient to
give a reasonable estimate of the value of the maximum halo energy. We adopt the
following procedure to determine  $\varepsilon_h$.
For a given initial distribution, we determine the maximum energy attained by a group of non-interacting test-spins
that are launched with the initial conditions selected from the distribution function, Eq.~(\ref{f0}). Their dynamical 
evolution  is governed by $\ddot\theta_i=-M_x(t)\sin\theta_i$  with
$M_x(t)$ determined by Eq.~(\ref{mx}), with $M_x(0)=M_0$ and $\dot M_x(t)=0$. We solve this equation 
over a short time corresponding to
two periods of oscillation of $M_x$. The $\varepsilon_h$, then, corresponds to the maximum energy obtained 
by any  of the test-spins.

Once  $\varepsilon_h$ has been determined using the test particle dynamics, 
the remaining parameters of the final stationary distribution --- $\varepsilon_F$,
$\chi$, and $M_s$ --- are obtained by imposing the conditions of conservation of norm  and of
the total energy, as well as the closure equation for magnetization:
\begin{eqnarray}
\int f_s (\theta,p)\, d\theta dp=1,\label{npa}\\
\int f_s (\theta,p)\, \varepsilon(\theta,p,M_s)\, d\theta dp=u,\label{ener}\\
\int f_s (\theta,p)\, \cos\theta\, d\theta dp=M_s.\label{ms}
\end{eqnarray}
The equations above can be analytically evaluated in terms of elliptic integrals, 
forming a closed set of algebraic equations that must be solved numerically to determine $\varepsilon_F$,
$\chi$, and $M_s$. 
%Integrating the distribution function over the phase space, 
%we can rewrite $\chi$ in terms of
%the other two parameters,
%\be
%\chi={4\theta_0p_0-\int \Theta(\varepsilon_F-\varepsilon) d\theta dp\over
%\int \Theta(\varepsilon_h-\varepsilon) \Theta(\varepsilon-\varepsilon_F) d\theta dp}.
%\ee
%Substituting this in eqs.~(\ref{ener}) and (\ref{ms}), we are left with only two algebraic equations that must be
%solved numerically to determine the Fermi energy and the final magnetization.

\begin{figure} 
\psfrag{E}{$\varepsilon$}
\includegraphics[scale=0.8,width=9cm]{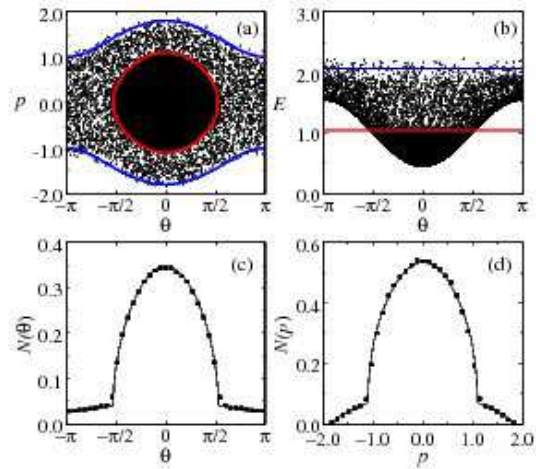}

\caption{Snapshots of (a) the phase-space and (b) of spin energies $\varepsilon$, as a function of angle.  The snapshot is taken 
at $t=10000$, using N=20000 spins. 
The solid curves correspond to the calculated Fermi energy $\varepsilon_F$ (red) and to the maximum halo energy $\varepsilon_h$ (blue), i.e.,
$\varepsilon(\theta,p,M_s)=p^2/2+1-M_s\cos\theta=\varepsilon_F$ and $\varepsilon(\theta,p,M_s)=\varepsilon_h$, respectively. 
We see that the Fermi energy curve perfectly encloses
the high density region.  The maximum halo energy obtained using the test particle dynamics and Eq.~(\ref{mx}) also delimits
well the extent of the particle distribution in the phase space (blue line). Panels (c) and (d) show the angle and the momentum
distributions, respectively. Solid curves are the theoretical predictions obtained using the distribution function
of Eq.~(\ref{fs}), and points are the results of molecular dynamics simulations averaged over 20 runs. 
The initial distribution has $M_0=0.80$ and $u=0.55$. Note that the present theory predicts that the 
maximum energy attained by any spin will be bounded by $\varepsilon_h$, while theories based on generalized entropies predict
that this energy distribution is unbounded, decaying either exponentially or algebraically~\cite{AnFa07b}, but see also~\cite{TePa09}.}
\label{fig1}
\end{figure}

\begin{figure}
\includegraphics*[scale=.45,clip=true,trim=.2cm 0 0 0]{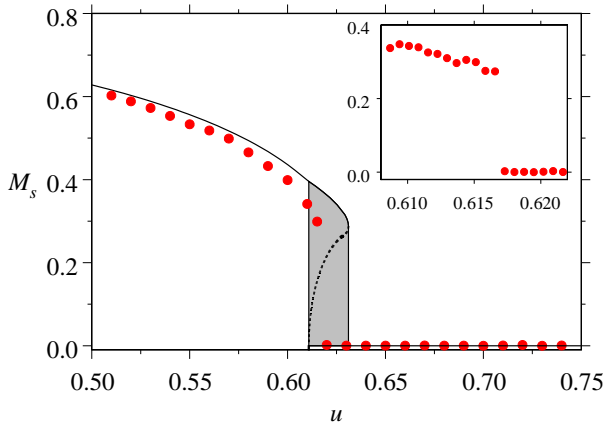}
\caption{Magnetization of the final stationary state $M_s$ as a function of the initial energy per particle $u$
for $M_0=0.40$. The curves are the theoretical predictions obtained using the distribution function
of Eq.~(\ref{fs}), and the points are the results of the molecular dynamics simulations with $N=10^6$ spins. 
The solid curve corresponds
to the stable solutions, whereas the dashed curve to unstable solutions. The grey area corresponds to the metastable
region in the parameter space where the phase transition must occur. The inset shows the result of 
a large number of molecular dynamics simulations, performed in the close 
vicinity of the phase transition. The abrupt change in magnetization,
as $u$ is varied indicates a first-order phase transition, as is predicted by the theory. For all simulations,
the integration was performed up to  $t=1200$.  The final  $M_s$ was obtained by averaging the magnetization
over an additional time interval $\Delta t=800$. We note that the theory based on Lynden-Bell's 
coarse-grained entropy incorrectly predicts that the phase transition at this point will be of second order~\cite{AnFa07}}
\label{fig2}
\end{figure}

\begin{figure}
\includegraphics*[scale=.45,clip=true,trim=.2cm 0 0 0]{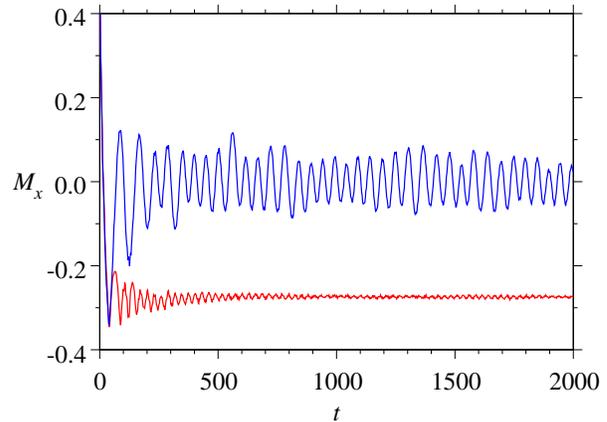}
\caption{Dynamical evolution of magnetization on two sides of the first-order phase-transition.  On the
paramagnetic side ($u=0.6170$), the magnetization oscillates around zero; while on the ferromagnetic side ($u=0.6165$), 
the oscillations are damped and magnetization converges to $M_s$ predicted by the theory.}
\label{fig3}
\end{figure}

In fig.(\ref{fig1}) (a and b) we show a snapshot of the phase-space  and of the distribution of spin energies.  
The core-halo separation can be seen very clearly. The energy $\varepsilon_h$ delimits 
the particle distribution over the phase space, while the Fermi energy restrict the extent of the core region.  In this
example the initial distribution has  $M_0=0.80$ and $u=0.55$, while the magnetization in the final stationary state is $M_s=0.56$.
In panels (c) and (d) we compare the theoretically calculated  spin and momentum 
distributions with the molecular dynamics simulations.
As can be seen, excellent agreement is found between the theory and the simulations.  

For all the initial distributions with finite magnetization, we find that if the energy per particle is less than $u<u_l$ there are 
two real roots of  eq. (\ref{ms}), $M_s=0$ and $M_s \ne 0$.  The root $M_s=0$ is unstable, so that only the 
solution with finite magnetization has a physical meaning.  On the other hand for $u>u_u$, there is only one real root, $M_s=0$.  For 
the interval of energies $u_l \le u \le u_u $ there are three distinct roots, see fig (\ref{fig2}).  The theory, therefore, 
predicts that there is a first order phase transition in the interval between $u_l \le u \le u_u $.  This is precisely
what is found in  simulations, see fig.(\ref{fig2}).  Unfortunately, differently from the equilibrium phase transitions, here we do not
have a free energy, which would allow us to precisely locate the transition point  using  
the Maxwell construction.  
All we can do is delimit the location of the first-order phase transition 
within a narrow interval $u_l \le u \le u_u $, shaded gray in  fig (\ref{fig2}). On the paramagnetic side of the phase
transition, the systems becomes trapped in an out-of-equilibrium limit cycle associated with appearance
of resonance islands in the phase space~\cite{BaCh08}, characterized by strong oscillation of magnetization
around $M=0$, see fig. (\ref{fig3}).

We have studied the paradigmatic system with long-range interactions the, so called, Hamiltonian Mean-Field Model. 
It is  shown that
in the thermodynamic limit this model does not relax to the usual Maxwell-Boltzmann distribution.  Instead the 
final stationary state of the HMF has a core-halo distribution function which is an ansatz solution to the Vlasov equation.
The theory allows us to explicitly calculate the spin and the momentum distribution functions, both of which 
are found to be in excellent
agreement with the simulations, {\it without any adjustable parameters}.  
We also find that the HMF model undergoes a first-order ferro-to-para phase transition in the region of parameter space
where earlier theories based on the Lynden-Bell (LB) coarse-grained entropy 
predicted a second order phase transition~\cite{AnFa07}. It is interesting to note that
the previous simulations~\cite{AnFa07,StCh09} --- performed over a fairly short time span $t<150$,  
before the system has fully relaxed --- failed
to notice this incorrect prediction of the LB theory.

The present
theory --- as well as the earlier results on non-neutral plasmas~\cite{Levinprl}, and self-gravitating systems 
in one~\cite{JoWo} and two~\cite{TaLe10} dimensions ---
suggests that there is a significant degree of universality in collisionless relaxation dynamics.
The core-halo distribution function appears to be a universal attractor  --- in a coarse-grained sense --- of systems with
long-range interactions, analogous to the MB distribution for collisional systems with short-range forces.

This work was partially supported by the CNPq, FAPERGS, INCT-FCx, and by the US-AFOSR under the grant FA9550-09-1-0283.
%\section*{}

\end{document}